\documentclass[%
 reprint,
 superscriptaddress,
 amsmath,amssymb,
 aps,
 prl,
]{revtex4-1}

\usepackage {atbegshi}
\usepackage[pdftex]{graphicx}
\usepackage{dcolumn}
\usepackage{bm}
\usepackage{braket}
\usepackage{multirow}
\usepackage {color}
\usepackage{txfonts}

\begin{document}
\newcommand\red[1]{\textcolor{red}{#1}}
\newcommand\blu[1]{\textcolor{blue}{#1}}

\preprint{APS/123-QED}

\title{
Universal voter model emergence in genetically labeled homeostatic tissues
}

\author{Hiroki Yamaguchi}
\email{yamaguchi@noneq.c.u-tokyo.ac.jp}
\affiliation{
 Department of Applied Physics, The University of Tokyo, 7-3-1 Hongo, Bunkyo-ku, Tokyo 113-8656, Japan
  }
\author{Kyogo Kawaguchi}
\email{kyogo.kawaguchi@riken.jp}
  \affiliation{
 Nonequilibrium physics of living matter research group, RIKEN BDR,  Kobe, Hyogo, Japan
}
\affiliation {
Universal Biology Institute, The University of Tokyo, 7-3-1 Hongo, Bunkyo-ku, Tokyo 113-0033, Japan
}

\date{\today}

\begin{abstract}
Recent experiments in adult mammalian tissues have found scaling relations of the voter model in the dynamics of the genetically labeled population of stem cells.
Yet, the reason for this seemingly robust appearance of the voter model remains unexplained.
Here we show that the voter model kinetics is indeed a generic behavior that arises at macroscale in a linearly stable homeostatic tissue undergoing turnover.
Starting from the continuum model of a multicellular system, we show that the dynamics of the labeled cell population converges to the voter model kinetics at large spatio-temporal scale of observation.
We present a method to calculate the length scale and time scale of coarse-graining that is required in obtaining the effective voter model dynamics, and apply it to the growth factor competition model and the pairwise mechanical interaction model.

\end{abstract}

\maketitle

\textit{Introduction.---}
Early models of stochastic interacting particle systems were introduced to study cell population competitions \cite{Eden1961, Williams1972} which led to general findings in nonequilibrium statistical physics including scaling relations and universal probability distributions in growing interfaces~\cite{Family1985, Kardar1986, Takeuchi2010, Takeuchi2011}.
Surface coarsening and growth dynamics in colony expansion and neutral drift dynamics in cultured cells and bacteria have been compared with statistical physics models~\cite{Matsushita1989, Huergo2012,Hallatschek2007, Korolev2010,McNally2017}.
In light of the recent developments in genetic engineering and imaging technologies, it is expected that more examples of model realizations will be discovered in real animal tissues, which will not only be intriguing from the physics viewpoint but also can be a useful paradigm for inferring parameters of cell kinetics to detect malignant conditions in our body.

An interesting experimental system where features of stochastic interacting particle systems have been observed is the homeostatic tissue~\cite{Klein2011}.
Adult cycling tissues are maintained through the dynamic balance of cell division, differentiation, and loss, as have been proven in mammals~\cite{Leblond1981,Blanpain2007} and flies~\cite{Ohlstein2006}.
Stem cells in tissues can be genetically labeled by drug induced methods and traced over time to measure the statistical properties of the labeled clone size, which is useful in inferring the rule of dynamcis in homeostasis~\cite{Klein2011, Simons2011}.
The seminiferous tubule~\cite{Klein2010, Hara2014}  and the intestinal crypt~\cite{Lopez-Garcia2010, Snippert2010}, two of the most classical examples in mammalian tissues that undergo rapid turnover~\cite{Leblond1952, Cheng1974} have shown the characteristics of the one-dimensional voter model~\cite{Klein2010, Hara2014, Snippert2010, Lopez-Garcia2010}.
The skin stem cells have also presented scaling results that are consistent with the two-dimensional voter model~\cite{Clayton2007,Mesa2018}.

The voter model is a well-studied model of an interacting stochastic process~\cite{Holley1975, Scheucher1988, Cox1986} which has been considered to describe the generic coarsening dynamics without surface tension~\cite{Dornic2001}.
The model predicts, for example, that the interface undergoes coarsening with slow logarithmic decay in a two-dimensional system, shown exactly for the lattice model \cite{Ben-Naim1996} and also in simulation studies using the continuum model \cite{Dornic2005,AlHammal2005}.
The voter model scaling found in the stem cell population of spermatogenesis~\cite{Klein2010, Hara2014} is particularly interesting since the cells are motile and sparsely distributed~\cite{Yoshida2018, Kitadate2019}, being far from the on-lattice setup of the original voter model~\cite{Holley1975}.
Appearance of voter model features at large length and time scales have been observed in numerical models~\cite{Yamaguchi2017,Jorg2019}.
However, it is still unclear why and under what conditions the voter model kinetics can appear in homeostatic tissues.

In this letter, we explain the experimental and numerical observations by showing that the voter model kinetics is indeed a robust property of the genetically labeled population in a homeostatic tissue.
That is, the field of labeled cell population denoted by $\varphi({\bf x })$ with the $(d+1)$-dimensional space-time coordinate $ {\bf x} := (\vec x; t)$, follows:
\begin {eqnarray}
	 \frac{\partial}{\partial t}{\varphi} ({\bf x })  = \tilde{D} \nabla^2_x \varphi ({\bf x })  + \sqrt{\tilde{\lambda} \varphi ({\bf x }) [1-\varphi ({\bf x }) ]  } \cdot \xi( {\bf x } ), \label{eq:voter}
\end {eqnarray}
under a well-defined coarse-graining, with the effective diffusion constant $\tilde{D}$ and turnover rate $\tilde{\lambda}$, the spatio-temporal white-Gaussian noise $\xi( {\bf x } )$ satisfying $\langle \xi( {\bf x }) \xi( {\bf x }') \rangle = \delta ({\bf x } - {\bf x }') $ with $\langle \cdot \rangle $ representing the statistical average, and  $\delta ({\bf x}) := \delta^d (\vec x) \delta (t)$. 
The multiplicative noise adopts the Ito convention.
Equation~(\ref{eq:voter}) exhibits key properties of the voter model~\cite{Dickman1995,Dornic2005, AlHammal2005}, and can be thought as the stochastic Fisher-Kolmogorov-Petrovskii-Piskunov equation under complete balance~\cite{Hallatschek2009}, as well as the compact directed percolation universality class at the critical point~\cite{Janssen2005}.

First we introduce a generic model of homeostatic cell density dynamics with a genetically labeled population. 
We then show how Eq.~(\ref{eq:voter}) can be obtained by coarse-graining the simplest model of linear density feedback, and consider how the required length and time scale of coarse-graining is determined in general models of homeostasis.
Using dynamical renormalization group, we show that the voter model kinetics is robust against details of the feedback dynamics such as length scales of feedback and mechanical interactions.
We analyze two important examples, the growth factor competition model and the mechanical interaction model, in order to see how the microscopic parameters of the chemical and cell kinetics appear or become irrelevant at the macroscopic level.
The results presented in this letter show how a model of nonequilibrium statistical physics emerges naturally under the simple assumption of tissue homeostasis.

\textit{Generic setup.---}
We start by modeling tissue stem cells as interacting particles labeled by $j=1,..., N(t)$ following the equations of motions in a $d$-dimensional space:
\begin {eqnarray}
	\frac {d}{dt} \vec x_j (t) = - \frac {\partial}{\partial \vec x_j } U (\{ \vec x \} ) + \sqrt {2D} \vec \eta_j (t) .
	\label{Eq:many-body}
\end {eqnarray}
Here, the potential term describes the pairwise mechanical interactions between cells: $ U ( \{ \vec x \} ) := \sum_{j=1}^{N(t)} \sum_{k < j} u (\vec x_j - \vec x_k) $, where $u (\vec x) $ describes the two-body interaction, and $D$ is the diffusion constant for a freely migrating cell. The components of the noise term $ \eta_j^{\mu} (t)$ are mutually independent white Gaussian with the correlation $\langle \eta_j^{\mu} (t) \eta_k^{\nu} (t^\prime) \rangle = \delta_{j,k} \delta^{\mu, \nu} \delta (t -t^\prime) $ for $\mu, \nu = 1, 2, ..., d$.
The number of cells $N(t)$ changes over time due to the stochastic birth-death process corresponding to cell division and elimination by differentiation or death, where ${\rm SC} \to {\rm SC} + {\rm SC}$ with rate $w^+$  and ${\rm SC} \to \emptyset$ with rate $w^-$. We assume that the birth-death rates depend on the position of the cell $\vec x_j (t) $ through spatiotemporal fields $\zeta ({\bf x}) $: the $j$-th cell follows the rates $w^{+/-}_j = w^{+/-} [\zeta(\vec x_ j, t)]$. The fields $ \zeta ({\bf x}) = \{ \zeta_i ({\bf x}) \}$ can include growth factor concentrations~\cite{Kitadate2019}, physical entities such as stress~\cite{Shraiman2005, Ranft2010}, the density of stem cells $\rho({\bf x}) := \sum_{j=1}^{N(t)} \delta(\vec{x} - \vec{x}_j (t))$ and the effects from other cells.
We neglect the $\zeta({\bf x})$-dependence of $U$ and $D$ for simplicity.

Starting from Eq.~(\ref{Eq:many-body}), the time evolution of the stem cell density $\rho ({\bf x })$ is obtained as~\cite{Dean1996, Yamaguchi2017, Supplementary}
\begin {eqnarray}
	 \frac{\partial}{\partial t}{ \rho} ({\bf x }) = D \nabla^2_x \rho ( {\bf x } ) + F[ \zeta ({\bf x}) ] + \xi_\rho ( {\bf x } ). \label{eq:genrho}
\end {eqnarray}
The second term, $F[\zeta ({\bf x}) ]  := \Delta w [\zeta ({\bf x}) ] \rho ({\bf x}) + \vec \nabla \cdot [ \rho ({\bf x})  (\vec \nabla u * \rho ) ({\bf x})  ]$, where  $\Delta w  [\zeta ({\bf x}) ] :=  w^+ [\zeta ({\bf x}) ]  - w^- [\zeta ({\bf x})] $ and $*$ denotes the convolution, arises from the birth-death kinetics and the cell-to-cell interactions (FIG.~1).
The noise term in Eq.~(\ref{eq:genrho}) satisfies	$\langle \xi_\rho ({\bf x} ) \xi_\rho ({\bf x }^\prime ) \rangle = \left ( g[\zeta({\bf x })] + 2 D \vec \nabla_x \cdot \vec \nabla_{x'} \right ) \rho ({\bf x } ) \delta ( {\bf x } - {\bf x }^\prime )$, where $ g[\zeta({\bf x })]:=  w^+ [\zeta ({\bf x}) ]  + w^- [\zeta ({\bf x})]$ describes the fluctuation of the birth-death process, and the second term corresponds to the conserved noise coming from diffusion.
The multiplicative noises are defined with the Ito convention throughout this letter.
For convenience, we set $\zeta_0({\bf x})=\rho ({\bf x })$.
The other components $\zeta_j ({\bf x})$ evolve, for example, according to the reaction-diffusion equation:
\begin {eqnarray}
	\frac {\partial}{\partial t} \zeta_{j} ({\bf x}) = \sigma_{j} \nabla^2 \zeta_{j} ({\bf x}) + F_{j} [ \zeta({\bf x}) ] +({\rm noise ~ term}),
	\label{Eq:reac-diff}
\end {eqnarray}
where $ \sigma_j$ is the diffusion constant and $F_j$ is the reaction term for the $j$-th field.

\begin{figure}[!t]
 \begin{center}
  \includegraphics[width=85mm]{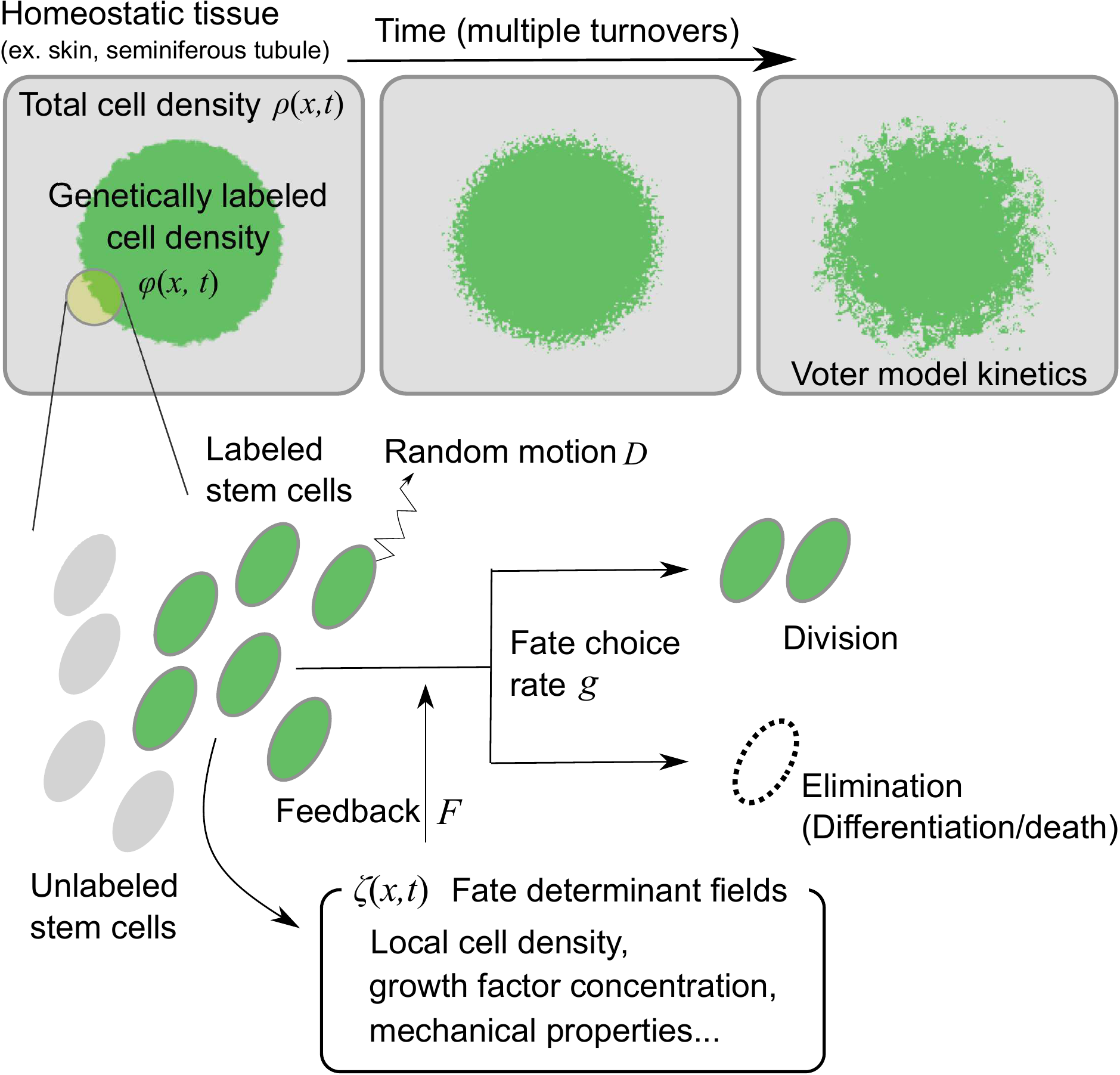}
 \end{center}
  \caption{\label{Fig1} (Color online) Tissue homeostasis for example in the skin and seminiferous tubule (spermatogenesis) is maintained while individual stem cells undergo rapid divisions and eliminations. Genetic labels (green) can be introduced to the cell populations by drug treatment or optogenetic methods to track the dynamics of their offsprings. We show that the macroscopic kinetics of the labeled cell population follows the voter model in the sense of Eq.~(\ref{eq:voter}).}
\end{figure}

Now, to model a homeostatic tissue, we restrict the whole dynamics so that there is a linearly stable steady-state solution $\{\zeta_{j, \rm ss}  \}=\{ \rho_{\rm ss}, \zeta_{1, \rm ss},... \}$ when neglecting the noise terms.
Then, we introduce a neutral genetic label to the subpopulation of the stem cells (FIG.~1) by setting $s_j = 0$ or 1, which is inherited to the daughter cells at every cell division.
The labeled cell density $ \varphi ({\bf x}) := \sum_{j=1}^{N(t)} s_j  \delta(\vec{x} - \vec{x}_j (t)) / \rho_{\rm ss}$ obeys the following equation~\cite{Supplementary}:
\begin {eqnarray}
	\frac {\partial}{\partial t} \varphi ({\bf x}) = D \nabla^2 \varphi ({\bf x}) +  F_\varphi[\zeta({\bf x}),\varphi({\bf x})] + \xi_\varphi ({\bf x}),
	\label{SupEq:lab-dens}
\end {eqnarray}
where $F_\varphi[\zeta({\bf x}),\varphi({\bf x})] := \{ w^+ [\zeta ({\bf x}) ]  - w^- [\zeta ({\bf x})]  \} \varphi ({\bf x}) + \vec \nabla \cdot [ \varphi ({\bf x})  ( \vec \nabla u * \rho  )  ] $.
Since the genetically labeled population is always a subset of the total population, $\xi_\varphi$ is correlated with $\xi_\mu$:
\begin {eqnarray} 
	\langle \xi_\varphi ({\bf x} ) \xi_\varphi ({\bf x }^\prime ) \rangle &=& \langle \xi_\varphi ({\bf x} ) \xi_\mu ({\bf x }^\prime ) \rangle   \nonumber\\
	&=& \left ( g[\zeta({\bf x})]+ 2 D \vec \nabla_x \cdot \vec \nabla_{x'} \right ) \varphi ({\bf x } ) \delta ( {\bf x } - {\bf x }^\prime ) / \rho_{\rm ss}. \label{eq:noiselabel}
\end {eqnarray}

Assuming that the fluctuation around the steady-state is small, we can linearize Eq.~(\ref{eq:genrho}) by introducing  $\epsilon ({\bf x})= \{\epsilon_{j} ({\bf x})   \} = \{\mu({\bf x}) , \epsilon_{1} ({\bf x}) ,... \} = \{[\rho({\bf x}) - \rho_{\rm ss}]/\rho_{\rm ss}, [\zeta_1({\bf x}) - \zeta_{1,\rm ss}]/\zeta_{1,\rm ss},... \}$. The time evolution of the cell density is now rewritten as
\begin {eqnarray}
	 \frac{\partial}{\partial t}{ \mu} ({\bf x}) &=&  D \nabla_x^2 \mu ({\bf x})  + {f} [\epsilon({\bf x} ) ] + \xi_{\mu} ({\bf x}). \label{eq:lineareq1}
\end {eqnarray}
Here, $f[\epsilon({\bf x})] :=\sum_{j \geq 0}  \delta \Delta w [\zeta] / \delta \zeta_j | _{\zeta = \zeta_{\rm ss} } \epsilon_j ({\bf x} ) + I_2({\bf x})$ with $I_2({\bf x}):= \int d \vec{y} [{\nabla}^2_x u (\vec{x}-\vec{y},t) ] \mu (\vec{y},t)$, and the noise term $\xi_{\mu}$ satisfies $\langle \xi_\mu ({\bf x} ) \xi_\mu ({\bf x }^\prime ) \rangle =  ( \lambda + 2 D \vec \nabla_x \cdot \vec \nabla_{x'}  ) \delta ( {\bf x } - {\bf x }^\prime ) / \rho_{\rm ss}$ where $\lambda:=g[\zeta_{\rm ss}]$ is the typical rate of the turnover.
We assume that in this regime the dynamics of $\zeta ({\bf x})$ also reduces to a linear equation for $\epsilon ({\bf x})$~\cite{Supplementary}.
The time evolution of ${\varphi} ({\bf x }) $ is now rewritten as
\begin {eqnarray}
  \frac{\partial}{\partial t}{\varphi} ({\bf x }) = D \nabla^2_x \varphi ( {\bf x } ) +  {f}_\varphi [ \epsilon({\bf x}), \varphi({\bf x })] + \xi_\varphi ( {\bf x } ). \label{eq:labeledcells}
\end {eqnarray}
with ${f}_\varphi [ \epsilon({\bf x}), \varphi({\bf x })] =f[\epsilon({\bf x})]  \varphi({\bf x}) + \vec{I}_1({\bf x}) \cdot \vec{\nabla} \varphi({\bf x})$ and $\vec{I}_1({\bf x}):= \int d \vec{y} [\vec{\nabla}_x u (\vec{x}-\vec{y},t) ] \mu (\vec{y},t)$.
Let us also  define $\eta_{\varphi}({\bf x}):=  {f}_\varphi [\epsilon({\bf x}), \varphi({\bf x })] + \xi_\varphi ( {\bf x } )$.

\textit{Density feedback model.---}
To see how the voter model dynamics [Eq.~(\ref{eq:voter})] arises from Eq.~(\ref{eq:labeledcells}), we first study the simplest case by setting $u (\vec x) =0$ and neglecting the effect of $\zeta_j ({\bf x})$ for $j \geq 1$ by setting ${f} [\epsilon({\bf x}) ] = - r \mu({\bf x}) $ with $r>0$.
In this model, the rates of the division and elimination of the cells are controlled by a linear feedback from the local cell density, i.e., the division rate is higher (lower) than the elimination rate when local density is low (high). 
By solving Eq.~(\ref{eq:lineareq1}), we obtain  $\langle \mu ({\bf x})  \mu ({\bf x}') \rangle  = ( \lambda+ 2 D \vec \nabla_x \cdot \vec \nabla_{x'}  )  M({\bf x-x}') / \rho_{\rm ss} r^2$ and $\langle \mu ({\bf x})  \xi_{\varphi} ({\bf x}') \rangle  = ( \lambda+ 2 D \vec \nabla_x \cdot \vec \nabla_{x'}  ) \varphi({\bf x}^\prime) N({\bf x-x}') / \rho_{\rm ss} r$, where $M$ and $N$ in  Fourier space can be written as
$\hat M (\vec k; \omega) = r^2/[ \omega^2 +  ( Dk^2 +r  )^2 ] $ and $\hat N (\vec k; \omega) =r/[ - i \omega+  Dk^2 + r  ]$ with the convention $ \hat \psi (\vec k; \omega) = \hat \psi ({\bf k}) = \int d^d \vec x \int d t  \psi ({\bf x}) e^{ - i \left ( \vec k \cdot \vec x - \omega t \right ) }$.
We introduce $\Delta ({\bf x} ):= N({\bf x} )+ N(-{\bf x} ) - M({\bf x} )$ and its Fourier tranform $\hat{\Delta} ({\bf k})$.

\begin{figure}[!t]
 \begin{center}
  \includegraphics[width=85mm]{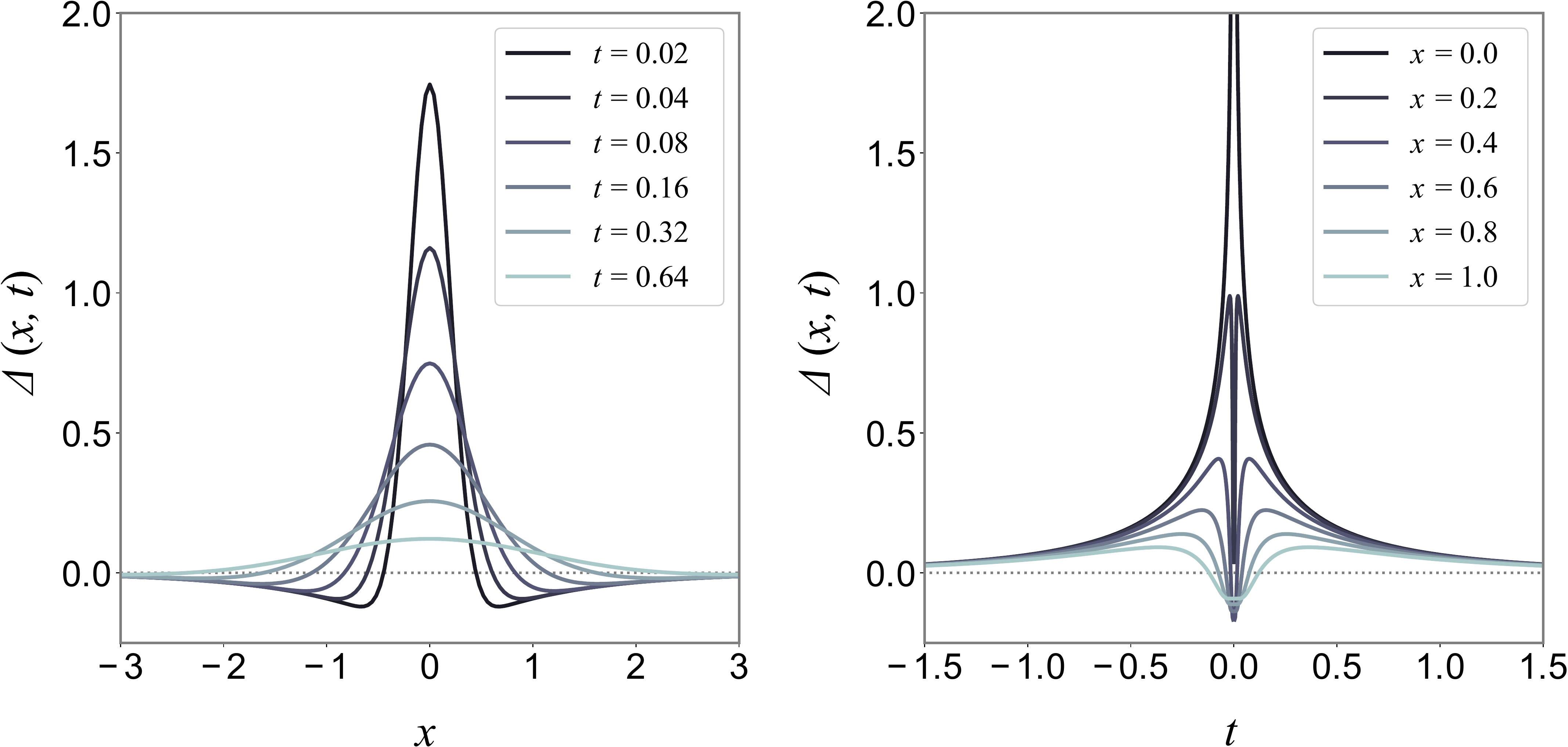}
 \end{center}
  \caption{\label{Fig2} (Color online) Space and time dependence of ${\Delta}({\bf x})$ which represents the contribution of the feedback term in the noise correlation [Eq.~(\ref{eq:etacorr})], with $D=r=1$. The  range of correlation is finite, meaning that  ${\Delta}({\bf x}) \sim 0$ for $|x| \gg {\sqrt{D/r} }$ and $|t| \gg 1/r$, indicating that ${\Delta}({\bf x})$ effectively becomes a delta function in a large enough spatio-temporal scale.  }
\end{figure}

We rewrite Eq.~(\ref{eq:labeledcells}) by introducing non-dimensional variables $\vec{X}= \vec{x}/L$ and $s=  t/\tau$, with $L$ and $\tau$ being the arbitrarily chosen units in length and time, respectively. Using ${\bf X}:= (\vec{X},s)$, we have
\begin {eqnarray} 
		 \frac{\partial}{\partial s }{  \varphi} ({\bf X })& =& \tilde{D} \nabla_X^2 \varphi ( {\bf X } )  + \tilde{\eta}_\varphi ( {\bf X } ),  \label{eq:newscale}
\end {eqnarray}
with
\begin {eqnarray}
 \langle \tilde{\eta}_ \varphi ( {\bf X } )\tilde{\eta}_\varphi ( {\bf X }' ) \rangle  &=&  (\tilde {\lambda} + 2\tilde{D}  \vec \nabla_X \cdot \vec \nabla_{X'}  )   \varphi ( {\bf X } ) [  \delta ( {\bf X } - {\bf X }^\prime )   \nonumber \\
		&&- \varphi ( {\bf X }^\prime ) \tilde{\Delta }_{L,\tau} ({\bf X } - {\bf X }^\prime) ] , \label{eq:etacorr}
\end {eqnarray}
where $\tilde{D} = D \tau/\tilde{\rho}_{\rm ss}L^2$, $\tilde{\lambda}=\lambda \tau/\tilde{\rho}_{\rm ss}$ with $\tilde{\rho}_{\rm ss}:={\rho}_{\rm ss} L^d$
, and $ \tilde{\Delta }_{L,\tau} ({\bf X } )$ is the inverse Fourier transform of $\hat{\Delta }  (\vec{k}L;  \omega\tau) $.
Figure \ref{Fig2} plots $\Delta({\bf x})$ in the case of $d=1$, $D=r=1$~\cite{Supplementary}.
Note that higher order cumulants of $\tilde{\eta}_\varphi$ are all zero since they can be rewritten in terms of the higher order cumulants of the white Gaussian noise terms.

In the limit of $L \gg L_D:=\sqrt{D/\lambda} $ and $\tau \gg {1/r}$, we find that $ \tilde{\Delta }_{L,\tau} ({\bf X } )$ converges to  $ \delta ({\bf X } )$~\cite{Supplementary}. In this limit, it also follows that $\tilde{D}/\tilde{\lambda}  \to 0$. Therefore, we obtain
\begin {eqnarray}
		 \langle \tilde{\eta}_\varphi ( {\bf X } )\tilde{\eta}_\varphi ( {\bf X }^\prime ) \rangle  \simeq \tilde{\lambda} \varphi ( {\bf X } ) [1 - \varphi ( {\bf X } )]   \delta ( {\bf X } - {\bf X }^\prime ).  \label{eq:etacorrlim}
\end {eqnarray}
Combining with Eq.~(\ref{eq:newscale}), we find that the dynamics of the density of the labeled fraction effectively follows the voter model  [Eq.~(\ref{eq:voter})] when observing the system at a larger length scale than $L_D$ and a longer time scale than $ {1/r}$.

\textit{Length and time scale of dynamics.---}
In the previous density feedback model, the dynamics of ${ \mu}({\bf x})$ had a finite time scale of $1/r$ whereas $\varphi ({\bf x })$ had no such scale since $-r \langle { \mu} ({\bf x})\rangle = 0$.
Thus, upon coarse-graining, the relatively fast dynamics of $\mu({\bf x})$ effectively turned into a Gaussian white noise that acts on the slow kinetics of $\varphi({\bf x})$. 

Here we see that the basic picture is the same for any linearly stable dynamics described by Eqs.~(\ref{eq:lineareq1},\ref{eq:labeledcells}).
For simplicity, we here neglect $u(\vec{x})$ and discuss its effect in the later example.
First, by solving the linear equations, we generally obtain $\hat{\mu} ({\bf k}) = \sum_{j \geq0} \hat{\xi}_{ j } ({\bf k})/\Lambda_{j}  ({\bf k})$,
where $\xi_{ j } $ is the noise terms affecting the dynamics of the $j$-th field.
Due to the linear stability condition (i.e., no neutral clustering~\cite{Young2001, Yamaguchi2017}), the factors $\Lambda_j ({\bf k})$ converge to non-zero constants, $a_j$, in the limit of small $k$ and $\omega$.
We can then introduce the characteristic length scale $L_\mu$ and time scale $\tau_\mu$ so that for all $j$, $\Lambda_j (\vec{k},\omega) \simeq a_j$ is satisfied for $k \ll 1/L_\mu$ and $\omega \ll 1/\tau_\mu$.

From Eq.~(\ref{eq:lineareq1}), $\hat{f}(\vec{k},\omega) = (-i \omega +Dk^2) \hat{\mu} (\vec{k},\omega) - \hat{\xi}_\mu (\vec{k},\omega)$.
Using $L_\mu$ and $\tau_\mu$, we have $\hat{f}(\vec{k},\omega) \simeq - \hat{\xi}_\mu (\vec{k},\omega)$ in the limit of  $k \ll 1/L_\mu$ and $\omega \ll 1/\tau_\mu$. This means that, when adopting the new scales $\vec{X}= \vec{x}/L$ and $s=  t/\tau$ with $L \gg L_\mu$ and $\tau \gg \tau_\mu$, we obtain
\begin {eqnarray}
\tilde{\eta}_{\varphi}({\bf X}) \simeq - \varphi({\bf X}) \xi_\mu ({\bf X}) + \xi_\varphi({\bf X}) . \label{eq:noiseapprox}
\end {eqnarray}
From Eq.~(\ref{eq:noiseapprox}) and further assuming $L \gg L_0 := \max \{L_D, L_\mu \}$, we can show Eq.~(\ref{eq:etacorrlim}).

The key property of the voter model kinetics is that the fluctuation is localized at the boundary, and there is no fluctuation in the bulk.
Noticing that $\xi_\varphi$ corresponds to the bulk fluctuation in the genetically labeled population, we find that the emergence of the voter model kinetics at large spatio-temporal scale owes to the suppression of the bulk noise by the feedback effect, which is the first term in Eq.~(\ref{eq:noiseapprox}).

The voter model kinetics is a robust feature of our generic model of our interest [Eqs.~(\ref{eq:lineareq1},\ref{eq:labeledcells})] also in the sense of universality and renormalization.
Firstly, the critical exponents of the voter model [Eq.~\eqref{eq:voter}] can be calculated using the symmetry upon transformation $\varphi ({\bf x}) \to 1 - \varphi ({\bf x}) $~\cite{Munoz1997, Supplementary}. For example, the exponent of the survival probability $P_{\rm surv} \sim t^{-\delta}$ is given by $\delta = d/2$ for $d\leq2$ (apart from the logarithmic correction for $d=2$), and $\delta = 1$ for $d > 2$.
This can be explained by the fact that the nontrivial fixed point in the renormalization flow, corresponding to the voter model, is infrared stable below the critical dimension $d_c=2$. The gradient terms and the finite range correlation terms in Eqs.~(\ref{eq:newscale}, \ref{eq:etacorr}), can be shown to be irrelevant by employing the one-loop calculation and the $\epsilon$-expansion~\cite{Wilson1974}. 
Therefore, for $d\leq2$, the flow of renormalization takes the system to the voter model fixed point, whereas for $d>2$, the same fixed point becomes unstable and the model essentially follows the mean-field dynamics described by the non-interacting critical birth-death kinetics~\cite{Supplementary}. 

\textit{Open niche competition model---}
From the previous analysis, we found not only that the voter model kinetics is a universal feature of homeostatic tissues, but also that the length and time scale of coarse-graining required to observe the voter model kinetics can be vastly different across tissues.
Here we will see how length scales other than $L_D$ can be involved in the feedback process by considering a minimal example where the external field plays a non-trivial role.

For the homeostasis of the seminiferous tubule, where the stem cells are motile and sparsely distributed along the surface of the tubule, it was recently reported that the abundance and exhaustion of an externally supplied growth factor promotes division and differentiation of stem cells, respectively~\cite{Kitadate2019}.
Assuming that the supply of the growth factor is constant and its decay depends on the uptake by the stem cells, there is an indirect feedback loop that stabilizes the stem cell density.
In the linearized dynamics~\cite{Jorg2019}, the stem cell density follows Eq.~\eqref{eq:lineareq1} with $f  =   r \chi ({\bf x}) $, where $\chi ({\bf x})$ is the linearized and normalized field of the growth factor, and $r>0$ is the coefficient that describes the promotion of stem cell division due to the growth factor. The growth factor field evolves according to 
\begin {eqnarray}
	\frac {\partial}{\partial t} \chi ({\bf x}) &=& - \beta  \chi ({\bf x}) - \kappa \mu ({\bf x}) + \sigma \nabla^2 \chi ({\bf x}) , \label{eq:chi}
\end {eqnarray}
where $\beta >0$ describes the natural decay, $\kappa>0$ represents the uptake rate by the stem cells, and $\sigma$ is the diffusion constant of this growth factor.
Under this condition, the dynamics is linearly stable, although clustering can still occur within the small fluctuation of the fields~\cite{Yamaguchi2017,Jorg2019}.

From Eqs.~(\ref{eq:lineareq1},\ref{eq:chi}), we obtain $\hat{\mu} (\vec{k},\omega) =  \hat{\xi}_{ \mu} (\vec{k},\omega)/\Lambda_{ \mu}  (\vec{k},\omega)$, where
\begin {eqnarray}
\Lambda_{ \mu}  (\vec{k},\omega) =  \alpha \left[ \frac{1}{1 -i \omega/\beta + \sigma k^2 /\beta } - i \frac{\omega}{\alpha} + \frac{D k^2}{\alpha} \right].
\end {eqnarray}
with $\alpha : = r \kappa/ \beta$.
The length scale and time scale in this system beyond which the voter model scaling should appear, are then given by 
\begin{eqnarray}
L_0 &=& \max \left \{ \sqrt{D/\alpha},\sqrt{\sigma/\beta} , L_D \right \} \label{eq:ONlength}
\end{eqnarray}
and
$\tau_\mu = \max \left \{ {1/\alpha}, {1/\beta} \right \}$.

Equation (\ref{eq:ONlength}) indicates that $L_0$ can be larger than $L_D$, for instance, when the diffusion of the growth factor is fast or the effective feedback time scale $1/\alpha$ is large.
In the case of the seminiferous tubule~\cite{Kitadate2019}, it is likely that the growth factor diffusion, represented by $\sigma$, is negligible since they are immobilized on the basement membrane.
According to the fit in \cite{Kitadate2019}, $\lambda$ is of the same order or larger than $\alpha$ and $\beta$, meaning that the largest relevant length scale is $\sqrt{D/\alpha}$, which is the typical length a cell travels within the time scale of the feedback.

\textit{Pairwise interaction model.---}
We consider the case where the cells are interacting with each other through a pairwise isotropic potential $u(\vec{x})$~\cite{Dean1996, Yamaguchi2017}.
A typical choice is the repulsive potential to describe volume exclusion with a length scale $L_c$ corresponding to the cell size and $\tau_c$ representing the typical time it takes for newborn sibling cells to relax to their positions.
Assuming additional density feedback with the time scale $1/r$, the linearized equations are obtained as
\begin {eqnarray}
	 	 \frac{\partial}{\partial t}{ \varphi} ({\bf x}) &=&  D \nabla_x^2 \varphi({\bf x})  -r \mu({\bf x}) \varphi({\bf x})  \nonumber \\
	 	 && + \vec{I}_1({\bf x}) \cdot \vec{\nabla}_{x} \varphi({\bf x})  + I_2({\bf x})  \varphi({\bf x})   + \xi_{\varphi}({\bf x}). \label{eq:pairwiselabel} \\
	 	 	 \frac{\partial}{\partial t}{ \mu} ({\bf x}) &=&  D \nabla_x^2 \mu ({\bf x})  -r \mu({\bf x}) + I_2({\bf x})   + \xi_{\mu}({\bf x}) \label{eq:pairwise} 
\end {eqnarray}
where the non-local effects due to cell-to-cell interactions are contained in  $\vec{I}_1({\bf x})$ and $I_2({\bf x})$.

The only difference from the previous examples is that there is an additional term that depends on $\vec \nabla_x \varphi$ in Eq.~(\ref{eq:pairwiselabel}).
First, from Eq.~(\ref{eq:pairwise}), we obtain $\Lambda_{ \mu}  ({\bf k}) = r [1 -i \omega/r + D k^2 /r - \hat{u}_2(k)/r]$, where $\hat{u}_2(k)$ is the Fourier transform of ${\nabla}^2_x u (\vec{x})$. 
Now, the Fourier transform of $\vec{I}_1({\bf x})$ can be written as ${\vec{I}}_1({\bf k})=\hat{\vec{u}}_1 ({\bf k}) \hat \xi_\mu({\bf k}) /\Lambda_{ \mu}  ({\bf k})$, with ${\vec{u}}_1 ({\bf k})$ being the Fourier transform of $\vec{\nabla}_x u (\vec{x})$.
We can then define $L_\mu:=\max \{ \sqrt {D/r}, L_c/{r \tau_c}, L_c/\sqrt{r \tau_c} \}$ and $\tau_\mu = 1/r$ so that $\Lambda_{ \mu}  (\vec{k},\omega) \simeq r$ and ${\vec{I}}_1({\bf k}) \simeq 0$ for $k \ll 1/L_\mu$ and $\omega \ll 1/\tau_\mu$.
Therefore, the characteristic length scale in the whole dynamics is $L_0 = \max  \{  \sqrt {D/r}, L_c/{r \tau_p}, L_c/\sqrt{r \tau_c} , L_D \}$.

We note that the scales of the mechanical interactions, $L_c$ and $\tau_c$, do not directly determine the scale of the fluctuation of $\mu$.
In fact, for $L_0$ to be finite, $r$ cannot be zero, which is the condition of linear stability.
For a simple critical birth-death process ($r=0$), largely interspaced cell clusters will grow at $t \to \infty$, even if pairwise repulsive interactions between the cells are working against the clustering.
As we saw in the open niche competition model, it is likely that the mechanism of stability is encoded directly or indirectly in the dynamics of the cell density in real tissues, effectively creating some time scale of feedback.

\textit {Discussion and conclusion.---}
Here we have shown that the genetically labeled population in a linearly stable cell density dynamics generally undergoes voter model kinetics in the coarse-grained spatio-temporal scale.
The suppression of the bulk fluctuation, which is the main character of the voter model, is achieved as a consequence of the feedback dynamics that stabilizes the homeostatic tissue, explaining how the voter model features appeared in the spermatogenic stem cell experiments~\cite{Klein2010, Hara2014}.

We saw how the length and time scale of coarse-graining required to observe the voter model depend on the nature of the feedback in the tissue.
This means that a multi-scale measurement of the genetically labeled population can be used to infer the relevant feedback scales.
Indeed, in the analysis of the live images of epidermal stem cells~\cite{Rompolas2016}, the fluctuation of the net cell imbalance was measured at various length and time scales in order to find that the feedback governing skin homeostasis is extremely short-ranged~\cite{Mesa2018}.
For fixed tissue experiments, it is also possible to measure the multi-scale statistical properties of genetically labeled populations, as have been demonstrated in the clonal labeling experiments~\cite{Clayton2007,Lopez-Garcia2010, Snippert2010,Klein2010}.
Our method bridges the gap between the statistical measurements provided in these examples to the microscopic parameters such as the mobility of the cells, the rate of growth factor uptake by the cells, and the mechanical properties.

Observing the coarsening of interfaces in real tissues will require larger scales both in space and time, which may be possible for example by analyzing somatic mutations in tissues~\cite{Martincorena2015, Simons2016}. 
The method we used to derive the effective kinetics of the labeled cell population can be extended to cases where there is heterogeneity in the population, for instance to model the initiation of cancer.
It is left for future work to analytically study the cases of wound healing and the taking over of malignant cells to find the general theory of cell population competitions in a multicellular system.

\begin{acknowledgments}
We are grateful to Allon Klein, Kazumasa Takeuchi, Takahiro Sagawa, and Yuki Minami for fruitful discussions and comments.
This work is supported in part by KAKENHI from Japan Society for the Promotion of Science (No. JP	18H04760, JP18K13515).
\end{acknowledgments}


\pagebreak

\onecolumngrid

\setcounter{equation}{0}
\setcounter{figure}{0}
\setcounter{table}{0}
\setcounter{page}{1}
\renewcommand{\theequation}{S\arabic{equation}}
\renewcommand{\thefigure}{S\arabic{figure}}
\renewcommand{\bibnumfmt}[1]{[S#1]}
\renewcommand{\citenumfont}[1]{S#1}

\begin {center} 
	\textbf{ \large Universal voter mode emergence in genetically labeled homeostatic tissues}\\ [.1cm]
	\textbf{ \large Supplemental material} \\ [.2cm]
	
	{  Hiroki Yamaguchi and Kyogo Kawaguchi}, \\ [.1cm]
	{(Dated: \today)} \\
\end {center}

\section {Derivation of continuum equations}
We here derive the continuum evolution equation for the cell density [Eqs.~(3,5,6) in the main text] from a general microscopic model combining the equations of motion [Eq.~(2)] and the stochastic cell division/differentiation kinetics
\begin {eqnarray}
	{\rm SC} \to& {\rm SC} + {\rm SC}  	&\hspace{0.2in} {\rm with~rate~} w^+, \\
	{\rm SC} \to& \emptyset		  	&\hspace{0.2in} {\rm with~rate~} w^-.
	\label{SupEq:birth-death}
\end {eqnarray}
Following the method presented by Dean~\cite{Sup:Dean1996}, the density of the $j$-th particle $\rho_j (\vec x, t) := \delta^d [\vec x_j(t) - \vec x ]$ follows the equation of motion with $(d+1)$-dimensional space-time coordinate ${\bf x}=(\vec x, t)$:
\begin {eqnarray}
	\frac {\partial}{\partial t} \rho_j ({\bf x}) = D \nabla^2 \rho_j ({\bf x}) + \vec \nabla \cdot \left [ \rho_j ({\bf x}) \left ( \vec \nabla u * \rho \right ) \right ] -  \vec \nabla \cdot \sqrt {2D \rho_j ({\bf x}) } \cdot \vec \eta_j (t) + B_j [\{\rho_j\}, \zeta] ,
	\label{SupEq:sing-part-dist}
\end {eqnarray}
where the multiplicative noise is defined with the Ito convention, and the convolution term is defined as
\begin {eqnarray}
	 (\vec \nabla u * \rho  ) ({\bf x}) := \int d\vec{y}  \rho (\vec{y},t) \nabla_x u(\vec{x} - \vec{y}).
	\label{SupEq:conv}
\end {eqnarray}
From the birth/death kinetics, we have the additional terms 
\begin {eqnarray}
	B_j [\{\rho_j\}, \zeta ] := \left \{ w^+ [\zeta ({\bf x}) ]  - w^- [\zeta ({\bf x})] \right \} \rho_j ({\bf x}) + \sqrt { \left \{ w^+ [\zeta ({\bf x}) ] + w^- [\zeta ({\bf x})] \right \} \rho_j ({\bf x}) } \cdot b_j ({\bf x}) ,
	\label{SupEq:birth-death-term}
\end {eqnarray}
where the noise term $b_j ({\bf x})$ is also white Gaussian with the correlation $\langle b_j ({\bf x}) b_k ({\bf x}^\prime) \rangle = \delta_{j,k} \delta ({\bf x} - {\bf x}^\prime)$, and $\zeta ({\bf x}) = \{ \zeta_i ({\bf x}) \} $ are the fields which evolve according Eq.~(4), for example.
Defining the total density $\rho ({\bf x}) := \sum_{j=1}^{N(t)} \rho_j ({\bf x} ) $, we obtain
\begin {eqnarray}
	\frac {\partial}{\partial t} \rho ({\bf x}) = D \nabla^2 \rho ({\bf x}) + F [ \zeta ({\bf x}) ] + \xi_\rho ({\bf x}) ,
	\label{SupEq:tot-dens}
\end {eqnarray}
where the feedback term is defined as
\begin {eqnarray}
	F[\zeta ({\bf x}) ]  = \left \{ w^+ [\zeta ({\bf x}) ]  - w^- [\zeta ({\bf x})] \right \} \rho ({\bf x}) + \vec \nabla \cdot \left [ \rho ({\bf x}) \left (\vec \nabla u * \rho \right ) ({\bf x}) \right ].
	\label{SupEq:feedback-term}
\end {eqnarray}
The noise term is given by
\begin {eqnarray}
	\xi_\rho ({\bf x}) := \sum_{j=1}^{N(t)} \left [ \sqrt { g [\zeta ({\bf x})] \rho_j ({\bf x}) } \cdot b_j (\vec x; t)  -  \vec \nabla \cdot \sqrt {2D \rho_j ({\bf x}) } \cdot \vec \eta_j (t) \right ],
	\label{SupEq:def-noise-rho}
\end {eqnarray}
which has the correlation
\begin {eqnarray}
	\langle \xi_\rho ({\bf x}) \xi_\rho ({\bf x}^\prime) \rangle &=&  \sum_{j, k=1}^{N(t)} \left \{ g[\zeta ({\bf x})] \rho_j ({\bf x}) \delta_{j,k} \delta ({\bf x} - {\bf x}^\prime ) + 2D \vec \nabla_x \cdot \vec \nabla_{x^\prime} \rho_j ({\bf x}) \delta_{j,k} \delta({\bf x} - {\bf x}^\prime) \right \} \nonumber \\
	&=& \left \{ g [\zeta({\bf x})] + 2D \vec \nabla_x \cdot \vec \nabla_{x^\prime} \right \} \rho ({\bf x}) \delta ({\bf x} - {\bf x}^\prime),
	\label{SupEq:corr-noise-rho}
\end {eqnarray}
where $g[\zeta ] := w^+ [\zeta] + w^- [\zeta] $.

We next introduce a genetic label to each cell, $ s_j = 0$ or 1, which is inherited to the daughter cells at every cell division. We assume that the kinetics of the cells does not depend on $s_j$. The labeled cell density $ \varphi ({\bf x}) := \sum_{j=1}^{N(t)} s_j \rho_j ({\bf x}) / \rho_{\rm ss}$ obeys the following equation of motion:
\begin {eqnarray}
	\frac {\partial}{\partial t} \varphi ({\bf x}) = D \nabla^2 \varphi ({\bf x}) +  \left \{ w^+ [\zeta ({\bf x}) ]  - w^- [\zeta ({\bf x})] \right \} \varphi ({\bf x}) + \vec \nabla \cdot \left [ \varphi ({\bf x}) \left ( \vec \nabla u * \rho \right ) \right ] + \xi_\varphi ({\bf x}).
	\label{SupEq:lab-dens}
\end {eqnarray}
We defined the noise term as
\begin {eqnarray}
	\xi_\varphi ({\bf x}) := \frac {1}{\rho_{\rm ss}} \sum_{j=1}^{N(t)} s_j \left [ \sqrt { \left \{ w^+ [\zeta ({\bf x}) ] + w^- [\zeta ({\bf x})] \right \} \rho_j ({\bf x}) } b_j (\vec x; t) -  \vec \nabla \cdot \sqrt {2D \rho_j ({\bf x}) } \vec \eta_j (t) \right ],
	\label{SupEq:def-noise-phi}
\end {eqnarray}
with the correlations
\begin {eqnarray}
	\langle \xi_\varphi ({\bf x}) \xi_\varphi ({\bf x}^\prime) \rangle &=& \frac {1}{\rho_{\rm ss}^2} \sum_{j, k=1}^{N(t)} s_j s_k  \left \{ g[\zeta ({\bf x})] \rho_j ({\bf x}) \delta_{j,k} \delta ({\bf x} - {\bf x}^\prime ) + 2D \vec \nabla_x \cdot \vec \nabla_{x^\prime} \rho_j ({\bf x}) \delta_{j,k} \delta({\bf x} - {\bf x}^\prime) \right \} \nonumber \\
	&=& \frac {1}{\rho_{\rm ss}} \left \{ g [\zeta({\bf x})] + 2D \vec \nabla_x \cdot \vec \nabla_{x^\prime} \right \} \varphi ({\bf x}) \delta ({\bf x} - {\bf x}^\prime), \\ 
	\langle \xi_\rho ({\bf x}) \xi_\varphi ({\bf x}^\prime) \rangle&=& \frac {1}{\rho_{\rm ss}^2} \sum_{j, k=1}^{N(t)} s_j  \left \{ g[\zeta ({\bf x})] \rho_j ({\bf x}) \delta_{j,k} \delta ({\bf x} - {\bf x}^\prime ) + 2D \vec \nabla_x \cdot \vec \nabla_{x^\prime} \rho_j ({\bf x}) \delta_{j,k} \delta({\bf x} - {\bf x}^\prime) \right \} \nonumber \\
	&=& \left \{ g [\zeta({\bf x})] + 2D \vec \nabla_x \cdot \vec \nabla_{x^\prime} \right \} \varphi ({\bf x}) \delta ({\bf x} - {\bf x}^\prime).
	\label{SupEq:corr-noise-phi}
\end {eqnarray}
Note that the cross-correlation does not vanish, since the labeled population is always a subset of the total population.

\section {Linearization}
We are interested in the homeostatic tissue, meaning that Eqs.~(4, \ref{SupEq:tot-dens}), without the noise terms, should have a stable steady-state solution, $\zeta_j ({\bf x}) = \zeta_{j, \rm ss} $. To see the condition for stability, we derive the linearized equations by introducing $ \epsilon_0 ({\bf x}) = \mu ({\bf x}) = (\rho ({\bf x}) - \rho_{\rm ss} )/\rho_{\rm ss} $ and $ \epsilon_j ({\bf x}) = ( \zeta_j ({\bf x}) - \zeta_{j, \rm ss}) / \zeta_{j, \rm ss} $ $(j = 1, 2, ...)$:
\begin {eqnarray}
	\frac {\partial}{\partial t} \epsilon_j ({\bf x}) &=& \sigma_j \nabla^2 \epsilon_j ({\bf x}) +   \sum_{k} f_{jl} \epsilon_l ({\bf x})  + \xi_j (t),
	\label{SupEq:linearization}
\end {eqnarray}
where $f_{jl} = \partial F_j[\zeta ({\bf x}) ]  / \partial {\zeta_l } |_{\zeta = \zeta_{\rm ss}}$ with $F_0[\zeta ({\bf x}) ]  = F[\zeta ({\bf x}) ] $ and $\sigma_0 = D$.
Note that 
\begin {eqnarray}
	\sum_j f_{0j} \epsilon_j ({\bf x})  = \sum_j \left. \frac{\delta}{\delta \zeta_j} \left \{ w^+ [ \zeta ({\bf x}) ]  - w^- [ \zeta ({\bf x}) ] \right \} \right |_{\zeta = \zeta_{\rm ss}}  \epsilon_j ({\bf x}) + ( \nabla_x^2 u * \mu ) ({\bf x}).
	\label{SupEq:linearf0}
\end {eqnarray}
Neglecting the noise terms in Eq.~(\ref{SupEq:linearization}) and considering the Fourier transform ($\vec{x} \to \vec{k}$), the linear stability condition is that the Jacobi matrix defined by $J_{jl} (\vec{k}) := - k^2 \sigma_i \delta_{jl} + f_{jl}(\vec k) $ is a stable matrix, i.e., the real parts of its eigenvalues are all negative.
The $\vec{k}$ dependence in $f_{jl}(\vec k)$ is due to the pairwise mechanical interactions.

With the noise terms, the linearized equations are solved in the Fourier space [${\bf x} \to {\bf k}:= (\vec{k},\omega)$] as
\begin {eqnarray}
	\hat \epsilon_j ({\bf k}) = \sum_{l} R_{jl} ({\bf k}) \hat \xi_l ({\bf k}) ,
\end {eqnarray}
where the response functions are given by 
\begin {eqnarray}
	R^{-1} _{ jl } ( {\bf k} ) := ( - i \omega + \sigma_j k^2 ) \delta_{ij} - f_{jl}(\vec k) .
\end {eqnarray}
Using this solution, we obtain
\begin {eqnarray}
	\hat f_j ({\bf k }) := \sum_l f_{jl} (\vec k) \hat \epsilon_{l} ({\bf k}) = \sum_{lm} f_{jl}  (\vec k) R_{lm} ({\bf k}) \hat \xi_m ({\bf k}) = \sum_m R^{f}_{jm} ({\bf k}) \hat \xi_m ({\bf k}).
\end {eqnarray}
Here, the length and time scales can be discussed using $R^{f}_{jm} ({\bf k}) := \sum_l f_{jl} (\vec k) R_{lm} ({\bf k}) $ as follows:
\begin{eqnarray}
	\left [ R^{f} \right ]^{-1}_{jl} = \sum_m \left [ ( -i \omega + \sigma_j k^2 ) \delta_{jm} - f_{jm} (\vec k) \right ] f^{-1}_{ml} (\vec k) = ( - i \omega + \sigma_j k^2 ) f^{-1}_{jl} (\vec k) - \delta_{jl} \simeq - \delta_{jl}.
\end{eqnarray}
That is, for sufficiently small ${\bf k}$, we can make $\omega \lvert f^{-1}_{jl} (\vec k) \rvert \ll 1$, and $ k \sqrt {\sigma_j  \lvert f^{-1}_{jl} (\vec k) \rvert } \ll 1 $. In particular, since we are interested in the dynamics of stem cell density $(j=0)$, we obtain
\begin{eqnarray}
	\hat f_0 ({\bf k}) = \sum_m R^f_{0m} ({\bf k}) \hat \xi_m({\bf k}) \simeq - \hat \xi_0 ({\bf k}).
\end{eqnarray}

\section {Derivation of noise correlation for the linear density feedback model}
In order to show the voter model dynamics in the long length and time scale, we here consider the simplest case of the linear density feedback model by setting $F [\zeta ({\bf x})] = r [ \rho_{\rm ss} - \rho ({\bf x}) ] \rho({\bf x}) $ with $r>0$, and omit the two-body potential term and the effect of $\zeta_j({\bf x})$ for $j \geq 1$. The linearized equations are given by
\begin {eqnarray}
	\frac {\partial}{\partial t} \mu ({\bf x}) &=& D \nabla^2 \mu ({\bf x}) - r \mu ({\bf x}) + \xi_\mu ({\bf x}), \label {SupEq:line-tot-dens} \\
	\frac {\partial}{\partial t} \varphi ({\bf x}) &=& D \nabla^2 \varphi ({\bf x}) - r \mu ({\bf x}) \varphi ({\bf x}) + \xi_\varphi ({\bf x}), \label{SupEq:line-lab-dens}	
\end {eqnarray}
where the noise term $\xi_\mu ({\bf x}) := \xi_\rho ({\bf x}) / \rho_{\rm ss} $ has the following correlations up to the leading order:
\begin {eqnarray}
	\langle \xi_\mu ({\bf x}) \xi_\mu ({\bf x}^\prime) \rangle &=&  \frac {1}{\rho_{\rm ss} } ( \lambda + 2D \vec \nabla_x \cdot \vec \nabla_{x^\prime} ) \delta ( {\bf x} - {\bf x}^\prime ) , \\
	\langle \xi_\mu ({\bf x}) \xi_\varphi ({\bf x}^\prime) \rangle	&=& \frac {1}{\rho_{\rm ss}} ( \lambda + 2D \vec \nabla_x \cdot \vec \nabla_{x^\prime} )\varphi ({\bf x}) \delta ({\bf x} - {\bf x}^\prime)
\end {eqnarray}
with $\lambda := g[\rho_{\rm ss}]$.

The fluctuation of the total stem cell density will behave like a white Gaussian noise, if we see the equation of motion for the clonal population density $\varphi ({\bf x})$ at a sufficiently long length and time scale.
Defining the effective noise $\eta ({\bf x}) := - r \mu ({\bf x}) \varphi ({\bf x}) + \xi_\varphi ({\bf x}) $, its correlation is given by
\begin {eqnarray}
	\langle \eta ({\bf x}) \eta ({\bf x}^\prime) \rangle =  \langle \xi_\varphi ({\bf x}) \xi_\varphi ({\bf x}^\prime) \rangle + r^2 \varphi ({\bf x}) \varphi ({\bf x}^\prime) \langle \mu ({\bf x}) \mu ({\bf x}^\prime) \rangle - r \varphi ({\bf x}) \langle \mu ({\bf x }) \xi_\varphi ({\bf x}^\prime) \rangle - r \varphi ({\bf x}^\prime ) \langle \xi_\varphi ({\bf x }) \mu ({\bf x}^\prime) \rangle .
\end {eqnarray}
By solving the linearized equation of motion [Eq.~\eqref{SupEq:line-tot-dens}] in the Fourier space, the density fluctuation is given by 
\begin {eqnarray}
	\hat \mu ({\bf k}) = \frac {1}{ - i \omega + Dk^2 + r } \hat \xi ({\bf k}), \label {SupEq:line-sol-tot}
\end {eqnarray}
with the convention $\hat \mu ({\bf k}) = \int d^d \vec x \int dt \mu ({\bf x} ) e^{ - i (\vec k \cdot \vec x - \omega t ) }$. The noise correlations in the Fourier space are given by
\begin {eqnarray}
	\langle \hat \xi_\mu ({\bf k}) \hat \xi_\mu ({\bf k}^\prime) \rangle &=& \frac {1}{\rho_{\rm ss} } (\lambda + 2 D k^2 ) (2 \pi)^{d+1} \delta ({\bf k} + {\bf k}^\prime), \label {SupEq:line-corr-noise-tot} \\
	\langle \hat \xi_\mu ({\bf k}) \hat \xi_\varphi ({\bf k}^\prime) \rangle &=& \frac {1}{\rho_{\rm ss} } (\lambda - 2 D \vec k \cdot \vec k^\prime )\varphi ({\bf k} + {\bf k}^\prime) \label {SupEq:line-corr-noise-cross}.
\end {eqnarray}
From Eqs.~(\ref{SupEq:line-sol-tot}, \ref{SupEq:line-corr-noise-tot}, \ref{SupEq:line-corr-noise-cross}), we calculate the correlations as follows
\begin {eqnarray}
	\langle \mu ({\bf x}) \mu ({\bf x}^\prime) \rangle &=& \int_{\bf k} \int_{{\bf k}^\prime }  \frac { \langle \xi_\mu ({\bf k}) \xi_\mu ({\bf k}^\prime) \rangle }{ (-i \omega + Dk^2 + r) ( - i\omega^\prime + Dk^{\prime 2} + r) } e^{i (\vec k \cdot \vec x - \omega t ) + i (\vec k^\prime \cdot \vec x^\prime - \omega^\prime t^\prime ) }\nonumber \\
	&=&  \int_{\bf k}  \frac { (\lambda + Dk^2) / \rho_{\rm ss} }{ \omega^2 + (Dk^2 + r)^2 } e^{i \vec k \cdot (\vec x - \vec x^\prime) -i  \omega (t - t^\prime) } = \int \frac {d^d \vec k}{(2 \pi)^d} \frac {(\lambda + 2Dk^2) / \rho_{\rm ss}}{2 (Dk^2 + r) } e^{ i \vec k \cdot \vec x -(Dk^2 + r )\lvert t \rvert } \nonumber \\
	&=& \frac {1}{\rho_{\rm ss}} (\lambda + 2D \vec \nabla_x \cdot \vec \nabla_{x^\prime} ) \frac {1}{r^2}M ({\bf x} - {\bf x}^\prime ), \\
	\langle \mu ({\bf x}) \xi_\varphi ({\bf x}^\prime) \rangle &=& \int_{\bf k} \int_{{\bf k}^\prime }  \frac { \langle \xi_\mu ({\bf k}) \xi_\varphi ({\bf k}^\prime) \rangle }{ -i \omega + Dk^2 + r } e^{i (\vec k \cdot \vec x - \omega t ) + i (\vec k^\prime \cdot \vec x^\prime - \omega^\prime t^\prime ) } =   \int_{\bf k} \int_{{\bf k}^\prime }  \frac { (\lambda - 2D \vec k \cdot \vec k^\prime ) / \rho_{\rm ss} }{ -i \omega^2 + Dk^2 + r } \hat \varphi (\hat k + \hat k^\prime ) e^{i (\vec k \cdot \vec x - \omega t ) + i (\vec k^\prime \cdot \vec x^\prime - \omega^\prime t^\prime ) } \nonumber \\
	&=& \frac {1}{\rho_{\rm ss}} (\lambda + 2D \vec \nabla_x \cdot \vec \nabla_{x^\prime} )\frac {1}{r} N ({\bf x} - {\bf x}^\prime ) \varphi ({\bf x}^\prime), 	
\end {eqnarray}
with the notation $\int_{\bf k}  := (2 \pi)^{-(d+1)} \int d^d\vec k \int d \omega $. The functions $M({\bf x})$ and $N({\bf x})$ are given by
\begin {eqnarray}
	M ({\bf x} ) &=& \frac {r}{2} e^{- r \lvert t  \rvert } \int d^d \vec y   \left ( \frac{1}{4 \pi D \lvert t \rvert } \right )^{d/2} \exp \left [ { - \frac{ (\vec x - \vec y)^2 }{ 4D \lvert t \rvert } } \right ]  (2 \pi)^{-d/2}  \lvert \vec y \rvert^{-\frac{d-2}{2}} K_{\frac {d-2}{2} } \left (\sqrt {r/D} \lvert \vec y \rvert \right ), \\
	N ({\bf x} ) &=& \Theta ( t ) r e^{ - r  t   } \left ( \frac{1}{4 \pi D  t  } \right )^{d/2}  \exp \left ( { - \frac{ \lvert \vec x \rvert ^2 }{ 4D  t  } } \right )  . 
\end {eqnarray}
where $K_{\nu} (z)$ is the $\nu$-th order modified Bessel function of the first kind, and $\Theta (t) $ is the step function. 

Taken together, the labeled population density follows
\begin {eqnarray}
	\frac {\partial}{\partial t} \varphi ({\bf x}) = D \nabla^2 \varphi ({\bf x}) + \eta ({\bf x}) \label{SupEq:voter-lab-dens}
\end {eqnarray}
with the noise correlation given by
\begin {eqnarray}
	\langle \eta ({\bf x}) \eta ({\bf x}^\prime) \rangle = \frac {1}{\rho_{\rm ss} } (\lambda + 2D \vec \nabla_x \cdot \vec \nabla_{x^\prime} ) \varphi ({\bf x} ) \left [ \delta ({\bf x} - {\bf x}^\prime ) - \varphi ({\bf x}^\prime ) \Delta ({\bf x} - {\bf x}^\prime ) \right ]. \label {SupEq:voter-lab-corr-noise}
\end {eqnarray}
where $\Delta ({\bf x} - {\bf x}^\prime ) = N ({\bf x} - {\bf x}^\prime ) + N ({\bf x}^\prime - {\bf x}) - M ({\bf x} - {\bf x}^\prime)$ is a normalized function, $ \int d{\bf x} \Delta ({\bf x}) =1$. If this correlation is approximately regarded as the delta function in a sufficiently long length and time scale, the dynamics will be governed by the voter model. In order to see how this works in the current example, we consider the long length and time scale dynamics by adopting the non-dimensionalization: $\vec X = x / L$, $s = t/\tau$, and ${\bf X} := (\vec X, s)$. Then, Eq.~\eqref{SupEq:voter-lab-dens} becomes
\begin {eqnarray}
	\frac {\partial}{\partial s} \varphi ({\bf X} ) = \tilde D \nabla^2_X \varphi ({\bf X} ) + \tilde \eta ({\bf X}), \label{SupEq:voter-lab-dens-resc}
\end {eqnarray}
where $\tilde D := D \tau / \tilde \rho_{\rm ss} L^2 $ using  $\tilde \rho_{\rm ss} := L^d \rho_{\rm ss}$, and 
\begin {eqnarray}
	\langle \tilde \eta ({\bf X}) \tilde \eta ({\bf X}^\prime) \rangle =( \tilde \lambda + 2 \tilde D \vec \nabla_X \cdot \vec \nabla_{X^\prime} ) \varphi ({\bf X} ) \left [ \delta ({\bf X} - {\bf X}^\prime ) - \varphi ({\bf X}^\prime ) \tilde \Delta ({\bf X} - {\bf X}^\prime ) \right ] 	\label {SupEq:voter-lab-corr-noise-resc}
\end {eqnarray}
with $\tilde \lambda := \lambda \tau/\tilde \rho_{\rm ss}$. 
To see how $\Delta ({\bf x} - {\bf x}^\prime)$ is transformed upon non-dimensionalization, we turn to the Fourier space, and obtain
\begin {eqnarray}
	 {\hat \Delta} ({\bf k} ) = 2 {\rm Re} [ \hat N ({\bf k}) ] - \hat M({\bf k}) = \frac { 1 + 2 Dk^2 / r  }{ \omega^2 / r^2 + (Dk^2 / r + 1)^2 } = \frac { 1 + \frac {2D}{r L^2} \lvert \vec K \rvert ^2 }{ \frac {\Omega^2 }{ (r \tau)^2} + \left ( \frac {D}{r L^2} \lvert \vec K \rvert ^2  + 1 \right ) ^2 } = \hat {\tilde \Delta} ({\bf K}) , 
\end {eqnarray}
where ${\bf K} := (\vec K, \Omega) = (\vec k L , \omega \tau ) $.
By taking the length and time scales as $L \gg \sqrt {D/r}$ and $\tau \gg 1/r$, we obtain $\hat {\tilde \Delta} ({\bf K}) \simeq 1$ for ${\bf K} = O(1)$, meaning that its inverse Fourier transform $\tilde \Delta ({\bf X} - {\bf X}^\prime )$ converges to the delta function from the viewpoint of ${\bf X} - {\bf X}^\prime \simeq O(1)$. Further noting that the gradient term in Eq.~\eqref {SupEq:voter-lab-corr-noise-resc} is negligible since $\tilde D / \tilde \lambda = D / \lambda L^2 \ll 1$ in this scale, Eq.~\eqref{SupEq:voter-lab-corr-noise-resc} becomes
\begin {eqnarray}
	\langle \eta ({\bf X}) \eta ({\bf X}^\prime) \rangle = \frac {\tilde \lambda}{ \tilde \rho_{\rm ss} } \varphi({\bf X}) [ 1 - \varphi({\bf x}) ] \delta ({\bf X} - {\bf X}^\prime ). \label{SupEq:voter-corr}
\end {eqnarray} 

Specifically, for $d = 1$, we obtain the following expression
\begin {eqnarray}
	 \Delta (x, t) = \frac {1}{ L_\mu \tau_\mu } \left \{ \frac { 1 }{\sqrt {4 \pi \lvert t \rvert / \tau_\mu } } \exp \left [{ - \frac { \left ( x / L_\mu \right )^2 }{ 4 \lvert t \rvert / \tau_\mu } - \frac {\lvert t \rvert }{ \tau_\mu } } \right ] - \frac{ 1 }{8} \exp \left ( { - \frac {x }{ L_\mu} } \right )  \left [ 2 - {\rm erfc} \left ( \frac { x / L_\mu - 2 \lvert t \rvert / \tau_\mu }{2 \sqrt { \lvert t \rvert / \tau_\mu } } \right )  + \exp \left ( { \frac { 2 x }{ L_\mu} } \right ) {\rm erfc} \left ( \frac {  x / L_\mu + 2 \lvert t \rvert / \tau_\mu }{2 \sqrt {  \lvert t \rvert / \tau_\mu } } \right )  \right ] \right \} , \nonumber \\
\end {eqnarray}
where $L_\mu := \sqrt {D/r}, \tau_\mu := 1/r$. We can directly check that $ \Delta (x, t) \simeq 0$ at $(x,t)$ satsiflying $|x| \gg L_\mu $ and $|t| \gg \tau_\mu $, as plotted in FIG.~2 of the main text. 

\section {Open niche competition model}
We consider the open niche competition model, where the fate decisions of stem cells are controlled by growth factors supplied from somatic cells in the external niche. Following the model presented in~\cite{Sup:Jorg2019}, we assume that the uptake of the molecule into stem cells promotes self-renewal and prevents differentiation. The molecules are supplied from outside of the system, decay spontaneously, consumed by stem cells, and can diffuse. 
Starting from Eqs.~(4, 5) in~\cite{Sup:Jorg2019}, expansion around the homeostatic steady state leads to the linearized equations:
\begin {eqnarray}
	\frac {\partial}{\partial t} \mu ({\bf x}) &=& D \nabla^2 \mu ({\bf x}) + r  \chi ({\bf x}) +  \xi_\mu ({\bf x}), \label {SupEq:line-dens-fgf} \\
	\frac {\partial}{\partial t}  \chi ({\bf x}) &=& \sigma \nabla^2 \chi ({\bf x}) - \beta \chi ({\bf x}) - \kappa \mu ({\bf x}), \label {SupEq:line-conc-fgf}
\end {eqnarray}
where $\chi ({\bf x})$ and $\mu ({\bf x})$ are the normalized concentration of fate determinant molecule and cell density, and the noise correlations are given by
\begin {eqnarray}
	\left < \xi_\mu ({\bf x}) \xi_\mu ({\bf x}^\prime) \right > = \frac{1}{\rho^*} \left ( \lambda + 2 D \vec \nabla_x \cdot \vec \nabla_{x^\prime} \right ) \delta ({\bf x} - {\bf x}^\prime).
\end {eqnarray}
By solving Eqs.~(\ref{SupEq:line-dens-fgf}, \ref{SupEq:line-conc-fgf}) in the Fourier space, we obtain the response functions as
\begin {eqnarray}
	R^f ({\bf k}) = \frac {-1}{ 1 + \alpha^{-1} (-i \omega + Dk^2 ) \left [1 + \beta^{-1} (-i\omega + \sigma k^2 ) \right ] }  \left [ 
	\begin{array}{cc}
		1 & \kappa^{-1} ( - i \omega + \sigma k^2 ) \\
		 r^{-1} ( - i \omega + D k^2 ) & 1 + \alpha^{-1} (-i \omega + D k^2)
	\end{array}
	\right ] .
\end {eqnarray}
with $\alpha := r \kappa / \beta$. Therefore, we obtain 
\begin{eqnarray}
	\hat f_0 ({\bf k}) = r \hat \chi ({\bf k}) = R^f_{00} ({\bf k}) \hat \xi_\mu ({\bf k}) = \frac {-1}{ 1 + \alpha^{-1} (-i \omega + Dk^2 ) \left [1 + \beta^{-1} (-i\omega + \sigma k^2 ) \right ] } \hat \xi_\mu ({\bf k}) , \label {SupEq:line-sol-fgf}
\end {eqnarray}
where the noise correlation is given by
\begin {eqnarray}
	\left < \hat \xi_\mu ({\bf k}) \hat \xi_\mu ({\bf k}^\prime) \right > = \frac{1}{\rho^*} \left ( \lambda + 2 D k^2 \right ) (2 \pi)^{d+1} \delta({\bf k} + {\bf k}^\prime). \label {SupEq:line-noise-corr-dens-fgf}
\end {eqnarray}

The labeled clonal population $\varphi ({\bf x}) $ follows the linearized equation
\begin {eqnarray}
	\frac {\partial}{\partial t} \varphi ({\bf x}) = D \nabla^2 \varphi ({\bf x}) + r  \chi ({\bf x}) \varphi ({\bf x}) +  \xi_\varphi ({\bf x}), 
\end {eqnarray}
where
\begin {eqnarray}
	\langle \xi_\mu ({\bf x}) \xi_\varphi ({\bf x}^\prime) \rangle =\langle \xi_\varphi ({\bf x}) \xi_\varphi ({\bf x}^\prime) \rangle = \frac{1}{\rho^*}(\lambda + 2 D \vec \nabla_x \cdot \vec \nabla_{x^\prime}) \varphi({\bf x}) \delta({\bf x} - {\bf x}^\prime ).  \label {SupEq:line-noise-corr-lab-fgf}
\end {eqnarray}

Using Eqs.~(\ref{SupEq:line-sol-fgf}, \ref{SupEq:line-noise-corr-dens-fgf}, \ref{SupEq:line-noise-corr-lab-fgf}), we can follow the same analysis as above, which yields 
\begin {eqnarray}
	\hat M ({\bf k} ) &=& \left \lvert \frac { r  \kappa  }{ ( -i \omega + \sigma k^2 + \beta ) ( -i \omega + Dk^2 ) + r  { \kappa } } \right \rvert^2 , \\
	\hat N ({\bf k}) &=& \frac { r  { \kappa } }{ ( -i \omega + \sigma k^2 + \beta ) ( -i \omega + Dk^2 ) + r  { \kappa } } ,
\end {eqnarray}
and therefore,
\begin {eqnarray}
	\hat \Delta (\hat k ) = 2 {\rm Re} [\hat N({\bf k} ) ] - \hat M({\bf k}) = \frac { 1 - 2 \omega^2 / \alpha \beta + 2 ( 1 + \sigma k^2 /\beta ) Dk^2 / \alpha }{ \left \lvert  1  + ( 1 - i \omega / \beta + \sigma k^2 / \beta ) ( -i \omega / \alpha + Dk^2 / \alpha ) \right \rvert ^2 } .\label{SupEq:Delta-corr-fgf}
\end {eqnarray}
Equation~\eqref{SupEq:Delta-corr-fgf} includes two different sets of length and time scales: the scale of the dynamics of the molecules $ \left (1 / \beta, \sqrt {\sigma / \beta} \right )$, and cells $ \left (1 / \alpha,  \sqrt {D / \alpha} \right )$. In non-dimensionalized coordinates, ${\bf K} = (\vec K, \Omega) = (\vec k L, \omega \tau)$, by setting $L \gg  \sqrt {D / \alpha},  \sqrt {\sigma / \beta} $ and $\tau \gg 1/\alpha, 1/\beta $, we obtain $\hat {\tilde \Delta } ({\bf K}) \simeq 1$ which means that the clonal labeled population effectively follows the voter model dynamics [Eqs.~(\ref{SupEq:voter-lab-dens-resc}, \ref{SupEq:voter-corr})].

\section {Renormalization group analysis}
We here perform the dynamical renormalization group analysis to show that Eqs.~(\ref{SupEq:voter-lab-dens-resc}, \ref{SupEq:voter-lab-corr-noise-resc}) at long length and time scales describe the universal behavior of the voter model. To this end, we turn to the field-theoretic description of the dynamics, following the Martin-Siggia-Rose-Janssen-de Dominicis formalism~\cite{Sup:Martin1973, Sup:Janssen1976, Sup:Dominicis1976, Sup:Dominicis1978}. First we define the probability density of $\varphi({\bf x})$ as  
\begin {eqnarray}
P[\varphi] &: =& \int D \tilde{\eta} P[\tilde{\eta}] \prod_ {\bf x} \delta \left [ \frac {\partial}{\partial t} \varphi ({\bf x} ) - \tilde D \nabla^2_X \varphi ({\bf x} ) - \tilde \eta ({\bf x})\right].
\end {eqnarray}
Here we use ${\bf x}=(\vec{x},t)$ instead of ${\bf X}$ for simplicity of notation. $P[\tilde{\eta}]$ denotes the path probabity of the noise term which respects the correlation [Eq.~(\ref{SupEq:voter-lab-corr-noise-resc})], and $\int D \tilde{\eta}$ is the functional integral.
By introducing the response field $\psi ({\bf x})$, we can further write
\begin {eqnarray}
P[\varphi] &=& \ \int D \left [ i \psi \right ]  D \tilde{\eta}  P[\tilde{\eta}] \exp \left \{ - \int d {\bf x} \psi ({\bf x}) \cdot \left[ \frac {\partial}{\partial t} \varphi ({\bf x} ) - \tilde D \nabla^2_x \varphi ({\bf x} ) - \tilde \eta ({\bf x}) \right] \right \}, \label{eq:pathpsi}
\end {eqnarray}
by using the formula for the delta function. The response field $\psi$ is integrated  along the imaginary axis, and
importantly, we adopt the Ito calculus for the product in the time integral so that the Jacobian terms become constant~\cite{Sup:Itami2017, Sup:Cugliandolo2017}. We can now integrate over $\tilde{\eta}$ to obtain
\begin {eqnarray}
P[\varphi] &=& \int D \left [ i \psi \right ]  \exp \left \{ -S [\varphi,  \psi] \right \} ,
\end {eqnarray}
where $S$ is the action functional:
\begin {eqnarray}
	S [\varphi,  \psi] :=  \int_{\bf k} \psi ( - {\bf k}) G_0^{-1} ({\bf k}) \varphi ({\bf k}) && {- \frac{1}{2} } \int_{{\bf k}_1, {\bf k}_2} \Gamma_0^{(1,2)} ({\bf k_1, }, {\bf k}_2 ) \psi ({\bf k}_1) \psi ({\bf k}_2) \varphi ( - {\bf k}_1 - {\bf k}_2 ) \nonumber \\
	&& {- \frac{1}{2} } \int_{{\bf k}_1, {\bf k}_2, {\bf k}_3 } \Gamma_0^{(2,2)} ({\bf k_1, k_2, k_3}) \psi ({\bf k}_1)  \psi ({\bf k}_2) \varphi ({\bf k}_3) \varphi ({\bf -k_1 - k_2 - k_3})  ,
\end {eqnarray}
defined with the bare propagator and the bare vertex functions:
\begin {eqnarray}
	G_0 ({\bf k} ) &=& \frac {1}{ -i \omega + D_0 k^2 }, \\
	\Gamma^{(1,2)}_0 ({\bf k_1, k_2} )&=& \lambda_1 - 2D_1 \vec k_1 \cdot \vec k_2,\\
	\Gamma^{ {(2,2)} }_0 ({\bf k_1, k_2, k_3} )&=& - \left (\lambda_2 - 2D_2 \vec k_1 \cdot \vec k_2 \right ) \frac {\Delta ({\bf k_1 + k_3}) + \Delta ({\bf k_2 + k_3}) }{2}.
\end {eqnarray}
Note that our starting point [Eqs.~(\ref{SupEq:voter-lab-dens-resc}, \ref{SupEq:voter-lab-corr-noise-resc})] has $ \lambda_1 = \lambda_2 = \tilde{\lambda}$, which is important for the $Z_2$ symmetry of the voter model. 
However, we here allow $\lambda_1 \neq \lambda_2$ to capture the general situations where $\lambda_2$ becomes irrelevant. 
We assume the lowest expansion of $\hat \Delta ({\bf k}) $:
\begin {eqnarray}
	\hat \Delta ({\hat k}) = 1 + L_0^2 k^2 + o(k^2).
\end {eqnarray}

We consider the bare parameter scalings upon transformation 
\begin{eqnarray}
	k  &\to& e^{-l} k, \nonumber \\
	\omega &\to& e^{-zl} \omega, \nonumber \\
	\varphi (\vec k, \omega) &\to& e^{\chi_\varphi l} \varphi (e^{-l} \vec k, e^{-zl} \omega ), \nonumber \\
	\psi (\vec k, \omega) &\to& e^{\chi_\psi l} \psi  (e^{-l} \vec k, e^{-zl} \omega ),
\end{eqnarray}
with $l > 0$. Since the action functional must be non-dimensional, we obtain the bare coupling scaling as
\begin{eqnarray} 
	D_0 &\to& e^{(z-2) l} D_0 ,\nonumber \\
	\lambda_1 &\to& e^{(\chi_\psi - d) l} \lambda_1 , \nonumber \\
	\lambda_2 &\to& e^{(z-d) l} \lambda_2  \nonumber \\
	D_1 &\to& e^{(\chi_\psi - d - 2 ) l} D_1 , \nonumber \\
	D_2 &\to& e^{ ( z - d - 2 ) l} D_2  \nonumber \\
	L_0^2 &\to& e^{-2l} L_0^2 ,
\end{eqnarray}
with $\chi_\varphi + \chi_\psi = d +2z $.

By demanding $D_0$ and $\lambda_1$ to be invariant upon scaling, we obtain the mean-field exponents 
\begin{eqnarray}
	z &=& 2, ~
	\chi_\psi = d,
\end{eqnarray}
and find that $\lambda_2, D_1$, $D_2$, and $L_0^2$ become irrelevant for $d>2$ since $\lambda_2 \to e^{(2-d)l} \lambda_2$, $D_1 \to e^{ -2 l }D_1$, $D_2 \to e^{ -d l} D_2 $, and $L_0^2 \to e^{-2l} L_0^2$.
Here, the mean-field dynamics is where the cells are undergoing non-interacting Brownian motion and the critical birth-death process (division and elimination).
The corresponding time evolution of the labeled cell density is
\begin {eqnarray}
	\frac{\partial}{\partial t} \varphi ({\bf x}) = D \nabla^2 \varphi ({\bf x}) + \sqrt { \lambda \varphi ({\bf x}) } \cdot \xi ({\bf x}) , \label{eq:meanfield}
\end {eqnarray}
with $ \langle \xi ({\bf x}) \xi ({\bf x}') \rangle = \delta({\bf x} -{\bf x}') $. In fact, the dynamics of Eq.~(\ref{eq:meanfield}) is known to display infinite 
 clustering for $d\leq 2$~\cite{Sup:Young2001}.

Below critical dimension $d < d_c = 2$, the four-point coupling $\lambda_2$ becomes relevant, and a nontrivial fixed point appears. In order to reveal this nontrivial fixed point, we perform the one-loop perturbation expansion to obtain the renormalization group (RG) equations. 
We introduce the generating functional 
\begin {eqnarray}
	Z [J_\varphi, J_\psi] = \int D \varphi D \left [i \psi \right ] \exp \left \{ - S [\varphi({\bf x}), \psi ({\bf x}) ] + \int d {\bf x} \left [ J_\varphi ({\bf x}) \varphi ({\bf x}) + J_\psi ({\bf x}) \psi ({\bf x})  \right ] \right \} 
\end {eqnarray}
and the vertex generating functional which is the Legendre transform of the generating functional~\cite{Sup:Tauber2014}
\begin {eqnarray}
	\Gamma [\Phi, \Psi] = - \ln Z [J_\varphi, J_\psi]  + \int d {\bf x} \left [ J_\varphi ({\bf x}) \Phi ({\bf x}) + J_\psi \Psi ({\bf x})  \right ],
\end {eqnarray}
with $\Phi ({\bf x} ) := \frac {\delta }{\delta J_\varphi ({\bf x})}  \ln Z [J_\varphi, J_\psi] , ~ \Psi ({\bf x} ) := \frac {\delta }{\delta J_\psi ({\bf x})}  \ln Z [J_\varphi, J_\psi] $.
The vertex functions are given by the derivatives 
\begin {eqnarray}
	\Gamma^{(N_\varphi, N_\psi)} ( \{ {\bf x}_m \} , \{ {\bf y}_n \} ) = \prod_{m=1}^{N_\varphi} \frac {\delta }{ \delta \Phi ( {\bf x}_m ) }  \prod_{n=1}^{N_\psi} \frac {\delta }{ \delta \Psi ( {\bf y}_n ) } \Gamma (\Phi, \Psi) |_{J_\varphi = J_\psi = 0}.
\end {eqnarray}

The one-loop corrections to the propagator and the vertex functions are given by
\begin {eqnarray}
	G^{-1}({\bf k}) &=& G _0^{-1} ({\bf k})  -  \int_{\bf q}  \Gamma^{(2, 2)}_0 ({\bf k}, {\bf q}, -{\bf q }) G_0 ({\bf q }), \label{eq:prop}\\
	\Gamma^{(1, 2) } ({\bf k_1,  k_2}) &=& 
\Gamma^{(1, 2)}_0 ({\bf k_1,  k_2}) +  2 \int_{\bf q} \Gamma_0^{(1, 2)} ({\bf q }, {\bf \bar k - q }) \Gamma^{(2, 2)}_0 ({\bf k}_1, {\bf k}_2, {\bf q }) G_0 ({\bf q }) G_0 ({\bf \bar k - q } ) \\
	\Gamma^{(2, 2) } ({\bf k_1,  k_2, k_3}) &=& \Gamma^{(2, 2)}_0 ({\bf k_1,  k_2, k_3}) + 2 \int_{\bf q} \Gamma^{(2, 2)}_0 ({\bf q }, {\bf \bar k - q }, {\bf k}_3) \Gamma^{(2, 2)}_0 ({\bf k}_1, {\bf k}_2, {\bf q }) G_0 ({\bf q }) G_0 ({\bf \bar k - q } ) , 
\end {eqnarray}
with ${\bf \bar k } := {\bf k}_1 + {\bf k}_2 $ and ${\bf q} = (\vec q, \Omega)$.
We find that the loop integral in Eq.~(\ref{eq:prop}) yields no constant term since~\cite{Sup:Tauber2014}
\begin {eqnarray}
\int _{\bf q}G_0 ({\bf q}) =  \int \frac {d^d \vec q}{(2 \pi)^d} G_0 (\vec q, t = 0) 
\end {eqnarray}
is zero due to the Ito convention we adopted for Eq.~(\ref{eq:pathpsi}).

In the following, we first perform the one-loop renormalization for $\lambda_2$ with $D_1, D_2, L_0^2 = 0$, and show that the fixed point indeed characterizes the voter model. Given the fixed point of the voter model, we then show that the other couplings $D_1, D_2, L_0^2$ do not affect the fixed point, in the sense of $\epsilon$-expansion~\cite{Sup:Wilson1974}.

We first evaluate the integral with vanishing external momentum/frequency: $k = 0, \omega = 0$. The frequency integral is performed by using the residue theorem. Then, we take the Wilsonian momentum shell approach~\cite{Sup:Wilson1974}, to gradually eliminate the short wavelength degrees of freedom. The momentum integral is evaluated within the momentum shell $q \in ( \Lambda e^{-l}, \Lambda ]$, where $l \ll 1$ and $\Lambda$ is the momentum cutoff. We then obtain the intermediate one-loop corrections as
\begin {eqnarray}
	\Gamma^{(1, 2) } ({\bf 0,  0}) &=& \lambda_1 - 2 \int_{\bf q} \frac {(\lambda_1 + 2D_1 \Lambda^2) \lambda_2  }{ \Omega^2 + ( D_0 q^2 )^2 } (1 + L_0^2 \Lambda^2 ) = \lambda_1 - l { \frac {K_d \Lambda^{d-2} }{ D_0 } } ( \lambda_1 + 2D_1 \Lambda^2 ) \lambda_2 (1 + L_0^2 \Lambda^2  ), \label {SupEq:1loop-corr-1} \\
	\Gamma^{(2, 2) } ({\bf 0, 0, 0}) &=& - \lambda_2 + 2 \int_{\bf q} \frac {(\lambda_2 + 2D_2 \Lambda^2) \lambda_2  }{ \Omega^2 + ( D_0 q^2 )^2 } (1 + L_0^2 \Lambda^2 )^2 = -\lambda_2 + l { \frac {K_d \Lambda^{d-2} }{ D_0 } } ( \lambda_2 + 2D_2 \Lambda^2 ) \lambda_2 (1 + L_0^2 \Lambda^2  )^2, 	\label {SupEq:1loop-corr-2}
\end {eqnarray}
where we define $K_d :=  (2 \pi)^{-d} S_d $, and $ S_d = 2 \pi^{d/2} / \Gamma (d/2)$ is the surface area of the $d$-dimensional sphere. After rescaling the momentum cutoff back to the original scale $ \Lambda e^{-l}  \to \Lambda$, we obtain the RG equations.

We here obtain the RG equations for $\lambda_1 , \lambda_2$, by first omitting $D_1, D_2, L_0^2$:
\begin {eqnarray}
	\frac {d}{dl} D_0 &=& (z-2) D_0, \\
	\frac {d}{dl} \lambda_1 &=& \left (  \chi_\psi - d  -  {\frac { K_d \Lambda^{d-2} }{D_0} } \lambda_2 \right ) \lambda_1, \\
	\frac {d}{dl} \lambda_2 &=& \left ( 2 - d - { \frac { K_d \Lambda^{d-2} }{D_0} } \lambda_2 \right ) \lambda_2 . \label{eq:RGlambda2}
\end {eqnarray}
Note that $D_0$ has no correction from the loop integral in Eq.~(\ref{eq:prop}). By introducing the non-dimensionalized coupling $ \gamma := { K_d \Lambda^{d-2} \lambda_2 / D_0 } $, Eq.~(\ref{eq:RGlambda2}) can be rewritten as
\begin {eqnarray}
	\frac {d}{dl} \gamma = \gamma ( 2-d - \gamma).
\end {eqnarray}
Apart from the trivial fixed point $\gamma = 0$ (mean-field), we find a nontrivial fixed point $\gamma = 2-d$, which becomes positive and stable for $d < 2$. In this nontrivial fixed point, $z = 2$ and  $\chi_\psi = 2$.
At each fixed point, the critical exponents are given by 
\begin{eqnarray}
	 -\beta / \nu_\perp &=& \chi_\varphi - d -z, \\
	  -\beta^\prime / \nu_\perp &=& \chi_\psi - d -z,
\end{eqnarray}
from which we can obtain the exponent of the survival probability, which is measurable in tissue experiments:
\begin{eqnarray}
	\delta = \beta^\prime / \nu_\parallel = -(\chi_\psi - d - z) / z .
\end{eqnarray}
Here, $ \nu_\perp, \nu_\parallel$ are the exponents of the correlation length and time near the critical point, respectively, and we used $z =  \nu_\parallel / \nu_\perp $. We obtain the mean-field exponent $\delta = 1$ for $d>2$, and the voter model exponent $\delta = d/2$ for $d \leq 2$ apart from the logarithmic correction for $d=2$ as shown in~\cite{Sup:Munoz1997, Sup:Duty2000}. 

Next, we see the effects of the couplings $D_1, D_2$ around the voter model fixed point. The one-loop correction to $D_1, D_2$ are generally given by 
\begin {eqnarray}
	\frac {d}{dl} D_1 &=& -d D_1 + I_1 \lambda_1 \lambda_2 + I_2 \lambda_1 D_2 + I_3 D_1 \lambda_2 + I_4 D_1 D_2 , \\
	\frac {d}{dl} D_2 &=& -d D_2 + I_5 \lambda_2^2 + I_6 \lambda_2 D_2 + I_7 D_2^2 ,
\end {eqnarray}
where $I_1, ..., I_7$ can be calculated by loop integrals, which we assume to have no singularity near $d = 2$. 
Introducing $\epsilon := 2-d$ as a small parameter, we find from Eq.~(\ref{eq:RGlambda2}) that $\lambda_2 = O(\epsilon)$ at the voter model fixed point.
Assuming $I_1, ..., I_7 = O (1)$, we also obtain $D_1 = O (\epsilon)$ and $D_2 = O (\epsilon^2)$ at the fixed point. We then turn to Eq.~\eqref{SupEq:1loop-corr-2} in order to take into account $D_1, D_2$, and obtain the RG equation for the effective coupling $\gamma$ as
\begin {eqnarray}
	\frac {d}{dl} \gamma = \gamma \left [ 2-d - \gamma \left ( 1 + \frac { 2D_2 \Lambda^2 }{\lambda_2} \right )  \right ].
\end {eqnarray}
Using $D_2 \Lambda^2 / \lambda_2 = O (\epsilon)$, we obtain $\gamma = \epsilon + O (\epsilon^2)$, meaning that the correction to the fixed point value by $D_2$ is negligible up to $O (\epsilon)$. From Eq.~\eqref{SupEq:1loop-corr-1}, we further obtain $\chi_\psi = d - \gamma (1+ 2D_1 \Lambda^2 / \lambda_1) = 2 + O (\epsilon^2)$, since $D_1 \Lambda^2 / \lambda_1 = O (\epsilon)$. Therefore, within one-loop order, the couplings $D_1, D_2$ do not affect the voter model fixed point for $d < 2$. Following the same argument to the coupling $L_0^2$, we can show that $L_0^2 \Lambda^2 = O (\epsilon)$, also cannot affect the voter model fixed point.

\end{document}